\documentclass[11pt]{article}

\usepackage{natbib}
\usepackage{amsfonts}
\usepackage{amsmath}
\usepackage{setspace}
\usepackage{graphicx}
\usepackage{verbatim}
\usepackage{color}
\usepackage{url}
\usepackage{endnotes}
\usepackage{paralist}
\usepackage{hyperref}

\newcommand*\chem[1]{\ensuremath{\mathrm{#1}}}

\begin{document}

\title{A Data-Driven Statistical Model for Predicting the Critical Temperature of a Superconductor }

\author{Kam Hamidieh \\ University of Pennsylvania, Wharton, Statistics Department}
\date{8/1/2018}
\maketitle

\begin{abstract}
We estimate a statistical model to predict the superconducting critical temperature based on the features extracted from the superconductor's chemical formula.  The statistical model gives reasonable out-of-sample predictions: $\pm 9.5$ K based on root-mean-squared-error.  Features extracted based on thermal conductivity, atomic radius, valence, electron affinity, and atomic mass contribute the most to the model's predictive accuracy. It is crucial to note that our model does not predict whether a material is a superconductor or not; it only gives predictions for superconductors. 
\end{abstract}


\section{Introduction}


Superconducting materials - materials that conduct current with zero resistance - have significant practical applications.  Perhaps the best known application is in the Magnetic Resonance Imaging (MRI) systems widely employed by health care professionals for detailed internal body imaging. Other prominent applications include the superconducting coils used to maintain high magnetic fields in the Large Hadron Collider at CERN, where the existence of Higgs Boson was recently confirmed, and the extremely sensitive magnetic field measuring devices called SQUIDs (Superconducting Quantum Interference Devices).  Furthermore, superconductors could revolutionize the energy industry as frictionless (zero resistance) superconducting wires and  electrical system may transport and deliver electricity  with  no energy loss; see \cite{Hassenzahl2000}.

However, the wide spread applications of superconductors have been held back by two major issues: (1) A superconductor conducts current with zero resistance only at or below its superconducting critical temperature ($T_c$).  Often impractically, a superconductor must be cooled to extremely low temperatures  near or below the boiling temperature of nitrogen (77 K) before exhibiting the zero resistance property.  (2) The scientific model and theory that \emph{predicts} $T_c$ is an open problem which has been baffling the scientific community since the discovery of superconductivity in 1911 by Heike Kamerlingh Onnes, in Leiden.

In the absence of any theory-based prediction models, simple empirical rules based on experimental results have guided researchers in synthesizing superconducting materials for many years.  For example, the eminent experimental physicist \cite{Matthias1955} concluded that $T_c$ is related to the number of available valence electrons per atom. (A few of these rules came to be known as the Matthias's rules.)    It is now well known that many of the simple empirical rules are violated; see \cite{second_life}.

In this study, we take an entirely data-driven approach to create a statistical model that predicts $T_c$ based on its chemical formula.  The superconductor data comes from the Superconducting Material Database maintained by Japan's National Institute for Materials Science (NIMS) at \url{http://supercon.nims.go.jp/index_en.html}.  After some data preprocessing, 21,263 superconductors are used.

To our knowledge, \cite{Stanev} and our work are the only papers that focus on statistical models to \emph{predict} $T_c$ for a \emph{broad class} of materials.  However, \cite{Owolabi2014} and \cite{Owolabi2015} focus on predicting $T_c$ for \chem{Fe} and \chem{MgB_{2}} based superconductors respectively.

We derive features (or predictors) based on the superconductor's elemental properties that could be helpful in predicting $T_c$.  For example, consider \chem{Nb_{0.8}Pd_{0.2}} with $T_c = 1.98$ K.  We can derive a feature based on the average thermal conductivities of the elements.  Niobium and  palladium's thermal conductivity coefficients are  54 and 71 W/(m$\times$K) respectively.  The mean thermal conductivity is $(54 + 71)/2 =  62.5$ W/(m$\times$K). We can treat  the mean thermal conductivity variable as a feature to predict $T_c$.  In total, we define and extract 81 features from each superconductor.

We tried various statistical models but we eventually settled on two: A multiple regression model which serves as a benchmark model, and a gradient boosted model as the main prediction model which is implemented in our  software.

Our software tool to predict $T_c$  and the associated data are available at \url{https://github.com/khamidieh/predict_tc} and will also be available at the publisher's complementary site.  We have done our best to make the software use and access to the data as easy as possible.

Gradient boosted models create an ensemble of trees to predict a response.   The trees are added in a sequential manner to  improve the model by accounting for the points which are difficult to predict. Once a gradient boosted model is fitted, the weighted average of all the trees is used to give a final prediction.  Gradient boosted models predict well because they are able to account for the complex interactions and correlations among the features.

The boosted models were first developed by \cite{schapire1990} and \cite{FREUND1995256}.  The boosted models were generalized to \emph{gradient} boosting by \cite{friedman2001}.  We use the latest improvement called XGBoost (eXtreme Gradient Boosting) by \cite{chen2016}, and the associated open-source R implementation of XGBoost by \cite{xgboost}.  XGBoost is also available in other popular programming languages such as python and Julia.  The full source code is at \url{https://github.com/dmlc/xgboost}.

Anthony Goldbloom, CEO of Kaggle (now a Google company), the premier data competition site,  stated: ``It used to be random forest that was the big winner, but over the last six months a new algorithm called XGBoost has cropped up, and it's winning practically every competition in the structured data category."  You can see the talk at \url{https://www.youtube.com/watch?v=GTs5ZQ6XwUM}.  Outside the competition realm, XGBoost has been successfully applied in disease prediction by \cite{chen2018}, and in quantitative structure
activity relationships studies by \cite{Sheridan2016}.

Our XGBoost model gives reasonable predictions: an out-of-sample error of  about $9.5$ K based on root-mean-squared-error (rmse), and an out-of-sample $R^2$ values of about $0.92$.  The numbers for the multiple regression model are about  $17.6$ K and $0.74$ for the out-of-sample rmse and $R^2$ respectively.  The multiple regression serves as a benchmark model.

We are able to assess the importance of the features in prediction accuracy. Features defined based on  thermal conductivity, atomic radius, valence,  electron affinity, and atomic mass are the most important features in predicting $T_c$.  On the downside, simple conclusions such as the exact nature of the relationship between the features and  $T_c$ can't be inferred from the XGBoost model.

\cite{Stanev} also create a model to predict $T_c$.  Our approach is different than \cite{Stanev} in the following ways: (1) We use XGBoost versus random forests, (2) we use a larger data set, (3) we use a single large model to obtain predictions rather than a cascade of models, (4) we create a larger number features \emph{only} from the elemental properties, and (5) most importantly, we quantify the out-of-sample prediction error.


\section{Data Preparation}

This section describes the detailed steps for the data preparation and feature extraction.  Subsection (\ref{subsec:Element_Data_Preparation}) describes how the element data is obtained and processed.  Subsection (\ref{subsec:Superconducting_Material_Data_Preparation}) describes the data preparation from NIMS Superconducting Material Database.  Subsection (\ref{subsec:Feature_Extraction}) details how the features are extracted.

\subsection{Element Data Preparation} \label{subsec:Element_Data_Preparation}
The element data with 46 variables and 86 rows (corresponding to 86 elements) are obtained by using the  \texttt{ElementData} function from Mathematica Version 11.1 by \cite{mathematica}.  Appendix (\ref{appendix:data_element}) lists the information sources for the element properties used by \texttt{ElementData}.  The first ionization energy data came from \url{http://www.ptable.com/} and is merged with the Mathematica data.  About 12\% of the entries out of the 3956 ($= 46 \times 86$) entries are missing.

In choosing the properties, we are guided by \cite{second_life} but we also use our judgement to pick certain properties.    For example, we drop the boiling point variable, and instead use the fusion heat variable which has no missing values, and is highly correlated with the boiling point variable.  We had also gained some experience and insight creating some initial models for predicting $T_c$ of elements only.  We settle on 8 properties shown in table (\ref{table:list_element_variables}).

\begin{table}
\scriptsize
\begin{center}
\begin{tabular}{|l|l|p{5cm}|}
  \hline
  \textbf{Variable} & \textbf{Units} & \textbf{Description} \\ \hline \hline
Atomic Mass                  & atomic mass units (AMU)                   & total proton and neutron rest masses   \\ \hline
First Ionization Energy      & kilo-Joules per mole (kJ/mol)             & energy required to remove a valence electron \\ \hline
Atomic Radius                & picometer (pm)                            & calculated atomic radius \\ \hline
Density                      & kilograms per meters cubed (kg/m$^3$)     & density at standard temperature and pressure \\ \hline
Electron Affinity            & kilo-Joules per mole (kJ/mol)             & energy required to add an electron to a neutral atom \\ \hline
Fusion Heat                  & kilo-Joules per mole (kJ/mol)             & energy  to change from solid to liquid without temperature change \\ \hline
Thermal Conductivity         & watts per meter-Kelvin (W/(m $\times$ K)) & thermal conductivity coefficient $\kappa$ \\ \hline
Valence                      & no units                                  & typical number of chemical bonds formed by the element \\ \hline
\end{tabular}
\caption{ This table shows the properties of an element which are used for creating features to predict $T_c$. \label{table:list_element_variables} }
\end{center}
\end{table}

With the choice of the above variables, we are only missing the atomic radii of La and Ce; we replace them with their covalent radii since atomic radii and covalent radii have very high correlation ($ \approx 0.95$) and approximately on the same scale and range.  Some bias may be introduced into our data with this minor imputation.   We add a small constant of 1.5 to the electron affinity values of all the elements to prevent issues when taking logarithm of $0$.

\newpage

\subsection{Superconducting Material Data Preparation} \label{subsec:Superconducting_Material_Data_Preparation}

Superconducting Material Database is supported by the NIMS, a public institution based in Japan.  The database contains a large list of superconductors, their critical temperatures, and the source references mostly from journal articles. To our knowledge, this is the most comprehensive database of superconductors.  Access to the database requires a login id and password but this is provided with a simple registration process.

We accessed the data on July 24, 2017 at \url{http://supercon.nims.go.jp/supercon/material_menu}.  Once logged in, we chose ``OXIDE \& METALLIC" material.  Figure (\ref{fig:menu}) shows a screen shot of the menu.  We clicked on the ``search" button to get \emph{all} the data.     We obtained 31,611 rows of data in a comma separated file format.  The key columns (variables) were ``element", the chemical formula of the material, and  ``Tc", the critical temperature.    Variable ``num" was a unique identifier for each row.  Column ``refno" contained links to the referenced source. The next few steps describe the manual clean up process:

\begin{figure}
\begin{center}
  \includegraphics[width=3.5in]{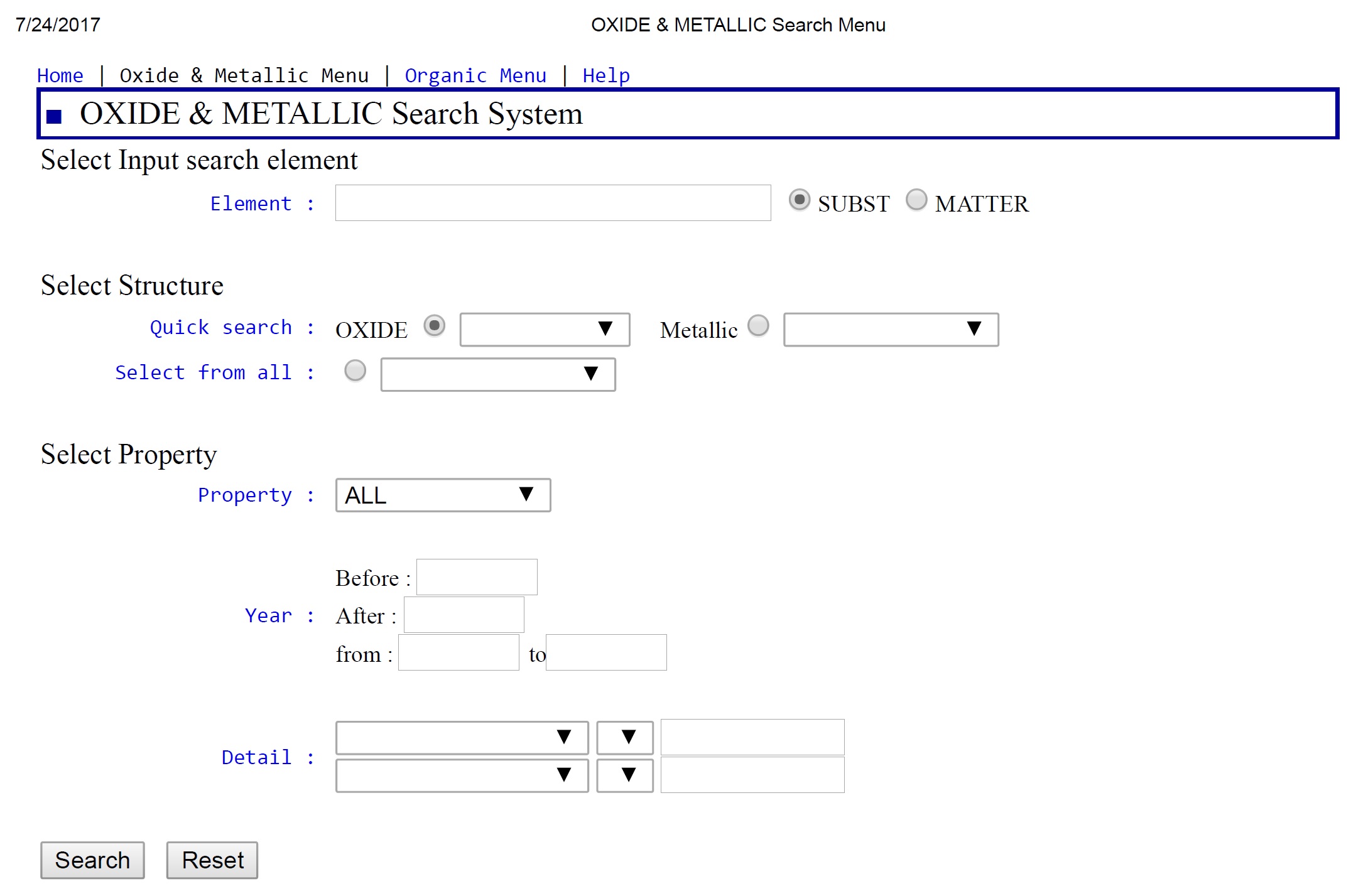}
  \caption{\small This is a screen shot of from Superconducting Material Database accessed on July 24, 2017. }\label{fig:menu}
  \end{center}
\end{figure}

\begin{figure}
\begin{center}
  \includegraphics[width=3.5in]{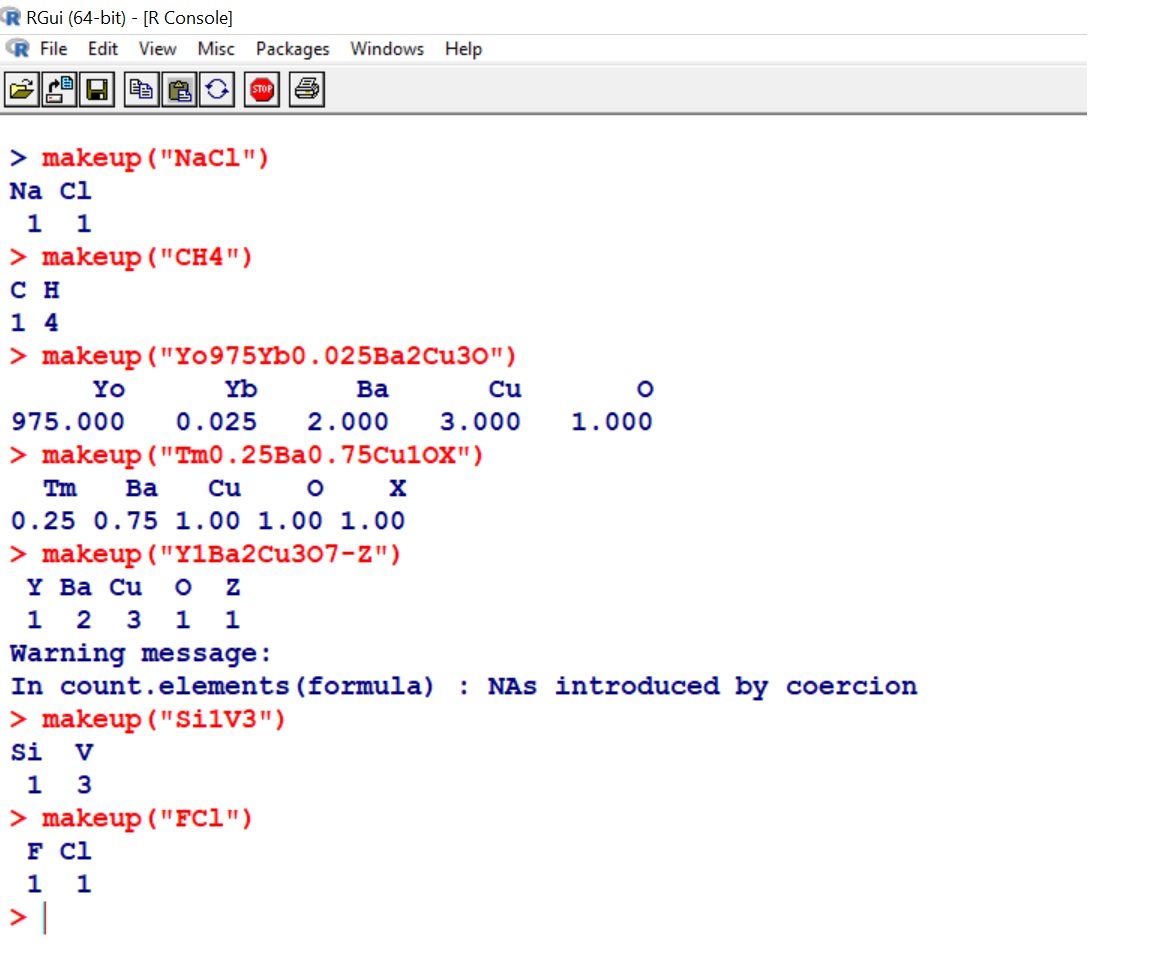}
  \caption{\small This screen shot is intended give you a sense of how the CHNOSZ package by \cite{CHNOSZ} works.  The first two materials \chem{NaCl} and \chem{CH4} are correctly broken up.  (These two are not superconductors and they are shown for illustration purposes.).  \chem{Yo_{975}Yb_{0.025}Ba_{2}Cu_{3}O} was a material in the database but this is obviously a mistake since no element with the symbol \chem{Yo} exists.  The same is true for the next material with \chem{X}.  However, no warnings are issued.  A warning is issued for \chem{Y_{1}Ba_{2}Cu_{3}O_{7-Z}}.  The next material \chem{SiV_{3}} was in the database and is correctly broken up.  \chem{FCl} is just given as another example.  It is not a superconductor and was not in the database.  The \texttt{makeup} command correctly breaks up the material but obviously does not check for the existence of \chem{FCl}. }\label{fig:makeup}
  \end{center}
\end{figure}

\begin{enumerate}
  \item We remove columns ``ma1" to ``mj2".
  \item We sort the data by ``Tc" from the highest to lowest.
  \item   The critical temperature for the following ``num" variables are mistakenly shifted by one column to the right. We fix these by recording them under the ``Tc" column: 31020, 31021, 31022, 31023, 31024, 31025, 153150, 153149, 42170, 42171, 30716, 30717, 30718, 30719,150001, 150002, 150003, 150004, 150005, 150006, 150007, 30712, 30713, 30714, 30715.
  \item The following are removed since the critical temperatures seemed to have been misrecorded; They have critical temperatures over 203 K which as of July 2017 was the highest reliable recorded critical temperature.
   La0.23Th0.77Pb3 (num = 111620), Pb2C1Ag2O6 (num = 9632), Er1Ba2Cu3O7-X (num = 140)	
   \item All rows with ``Tc" = 0 or missing are removed.
   \item Columns with headings ``nums", ``mo1", ``mo2", ``oz", ``str3", ``tcn", ``tcfig", ``refno" are removed.
   \item We manually chang all materials with oxygen content formula such as O7-X to the best oxygen content approximation.  For example, O7-X is changed to O7, O5+X is changed to O5, etc.  This certainly introduces some error into our data but it is impossible to go document by document  to get better estimates of the oxygen contents.  At this point our data has two columns: ``element" and ``Tc".
   \item We use  R statistical software by \cite{r} and the CHNOSZ package by \cite{CHNOSZ} to perform a preliminary check of the validity of the chemical formulas.  The CHNOSZ package has a function \texttt{makeup} which reads the chemical formula in string format and breaks up the formula into the elements and their ratios.  In some cases, it throws an error or a warning when the chemical formula does not make sense.  For example it throws a warning message if Pb-2O is checked; Negative number of Pb does not make sense.  However, the function does not check whether the material could actually exist.  See figure (\ref{fig:makeup}) to get a sense of how this function works.  With the help of the CHNOSZ package, we make the following modifications:
       \begin{enumerate}
         \item Yo975Yb0.025Ba2Cu3O, Yo975Yb0.025Ba2Cu3O, Yo975Yb0.025Ba2Cu3O are removed.  There is no element with the symbol Yo.  It's likely that Y0.975 was misrecorded as Yo975 but we can't be sure.
         \item Bi1.7Pb0.3Sr2Ca1Cu2O0, La1.85Nd0Ca1.15Cu2O5.99, Bi0Mo0.33Cu2.67Sr2Y1O7.41, \\ Y0.5Yb0.5Ba2Sr0Cu3O7 are removed since some elements had coefficients of zero.
         \item Y2C2Br0.5!1.5 is removed.  The exclamation sign throws an error message.
         \item Y1Ba2Cu3O6050 is removed.  The coefficient of 6050 for oxygen is possibly a mistake.
         \item Hg1234O10 is removed.  The coefficient of 1234 for mercury is possibly a mistake.
         \item Nd185Ce0.15Cu1O4 is removed.  The coefficient of 185 for Neodymium  is possibly a mistake. There is a Nd1.85Ce0.15Cu1O4 already in the data.
         \item Bi1.6Pb0.4Sr2Cu3Ca2O1013 is changed to Bi1.6Pb0.4Sr2Cu3Ca2O10.13 since nearby rows in the data have formulas with O10.xx.
         \item Y1Ba2Cu285Ni0.15O7 is changed to Y1Ba2Cu2.85Ni0.15O7 since nearby rows in the data have formulas with Cu2.xx.
       \end{enumerate}
    \item The column headings of ``Tc" and ``element" are changed to ``critical\_temp" and ``material" respectively.
\end{enumerate}

6750 rows are left out because $T_c$ is either zero or missing. At this point we have 24,861 rows of data.

The rest of the data preparation is done in \cite{r}.  We exclude any superconductor that has an element with an atomic number greater than 86.  This eliminates an additional 973 rows of data.  For example, superconductors that have uranium are left out.  We remove the repeating rows.  It would be impossible to manually check to see whether the repeated rows are genuine independent reports from independent experiments or they are just duplicate reportings.  After all the data preparation and clean up, we end up with 21,263 rows of data or about 67\% of the original data we started with.

\newpage
\subsection{Feature Extraction} \label{subsec:Feature_Extraction}

In this section, we describe the feature extraction process through a detailed example: Consider \chem{Re_{7}Zr_{1}} with $ T_c = 6.7$ K, and focus on the features extracted based on thermal conductivity.

Rhenium and  Zirconium's thermal conductivity coefficients are  $t_1 = 48$ and $t_2 = 23$ W/(m$\times$K) respectively.  The ratios of the elements in the material are used to define features:
\begin{equation} \label{equ:ratios}
p_1 = \frac{6}{6+1}= \frac{6}{7}, \ \ \  p_2 = \frac{1}{6+1}= \frac{1}{7}.
\end{equation}
The fractions of total thermal conductivities are used as well:
\begin{equation} \label{equ:feature_value_ratios}
w_1 = \frac{t_1}{t_1+t_2} = \frac{48}{48+23}= \frac{48}{71}, \ \ \ w_2 = \frac{t_2}{t_1+t_2} = \frac{23}{48+23}= \frac{23}{71}.
\end{equation}
We need a couple of intermediate values based on equations (\ref{equ:ratios}) and (\ref{equ:feature_value_ratios}):
$$ A = \frac{p_1w_1}{p_1w_1 + p_2w_2} \approx 0.926, \ \ \ B = \frac{p_2w_2}{p_1w_1 + p_2w_2} \approx 0.074.$$

Once we have  obtained the values $p_1, p_2, w_1, w_2, A,$ and $B$, we can extract 10 features from Rhenium and  Zirconium's thermal conductivities as shown in table  (\ref{table:example_feature_table}).

\begin{table}
\footnotesize
\begin{center}
\begin{tabular}{|l|l|l|}
  \hline
  \textbf{Feature \& Description} & \textbf{Formula} & \textbf{Sample Value} \\ \hline
  Mean                        & $ = \mu = (t_1 + t_2)/2 $                                      & $35.5  $ \\  \hline
  Weighted mean               & $ = \nu = (p_1  t_1) + (p_2  t_2) $                & $44.43 $ \\  \hline
  Geometric mean              & $ = (t_1  t_2)^{1/2} $                             & $33.23 $ \\  \hline
  Weighted geometric mean     & $ = (t_1)^{p_1}  (t_2)^{p_2} $                     & $43.21 $ \\  \hline
  Entropy                     & $ =  -w_1  \ln(w_1) - w_2  \ln(w_2) $       & $0.63  $ \\  \hline
  Weighted entropy            & $ = -A\ln(A) - B\ln(B) $       & $0.26  $ \\  \hline
  Range                       & $ = t_1  - t_2  \ ( t_1  > t_2) $       & $25 $ \\  \hline
  Weighted range              & $ = p_1 t_1 - p_2 t_2 $       & $37.86 $ \\  \hline
  Standard deviation          & $ = [(1/2)((t_1-\mu)^2 + (t_2-\mu)^2)]^{1/2}$       & $12.5 $ \\  \hline
  Weighted standard deviation & $ = [p_1(t_1-\nu)^2 + p_2(t_2-\nu)^2)]^{1/2}$       & $8.75$ \\  \hline
\end{tabular}
\caption{ \small This table summarizes the procedure for feature extraction from material's chemical formula. The last column serves as an example;  features based on thermal conductivities for  \chem{Re_{7}Zr_{1}}are derived and reported to two decimal places. Rhenium and  Zirconium's thermal conductivity coefficients are  $t_1 = 48$ and $t_2 = 23$ W/(m$\times$K) respectively.  Here: $p_1 = \frac{6}{7}, p_2 = \frac{1}{7}, w_1 = \frac{48}{71}, w_2 = \frac{23}{71}, A = \frac{p_1w_1}{p_1w_1 + p_2w_2} \approx 0.926, B = \frac{p_2w_2}{p_1w_1 + p_2w_2} \approx 0.074.$\label{table:example_feature_table} }
\end{center}
\end{table}

We repeat the same process above with the 8  variables listed in table (\ref{table:list_element_variables}).  For example, for features based on atomic mass, just replace $t_1$ and $t_2$ with the atomic masses of Rhenium and  Zirconium respectively, then carry on with the calculations of $p_1, p_2, w_1, w_2, A, B$, and  finally  calculate the 10 features defined in table (\ref{table:example_feature_table}).  This gives us $8 \times 10 = 80$ features.  One additional features, a numeric variable counting the number of elements in the supercondutor, is also extracted.  We end up with 81 features in total.

In summary: We have data with 21,263 rows and 82 columns: 81 columns corresponding to the features extracted and 1 column of the observed $T_c$ values.

We also considered but did not implement features that simply indicate whether an element is present in the superconductor or not. For example, we could have had a column that indicated whether say oxygen is in the material or not.  However, this approach would have added a large number of indicator variables to our data, made model selection and assessment too complicated, and increased the chances of over-fitting.


\section{Analysis}

This section has two parts: Basic summaries of the data are given in subsection (\ref{subsect:descriptive_analysis}).  The statistical models are described in subsection (\ref{subsect:model_analysis}).

\subsection{Descriptive Analysis}  \label{subsect:descriptive_analysis}

\begin{figure}
\begin{center}
  \includegraphics[width=6in]{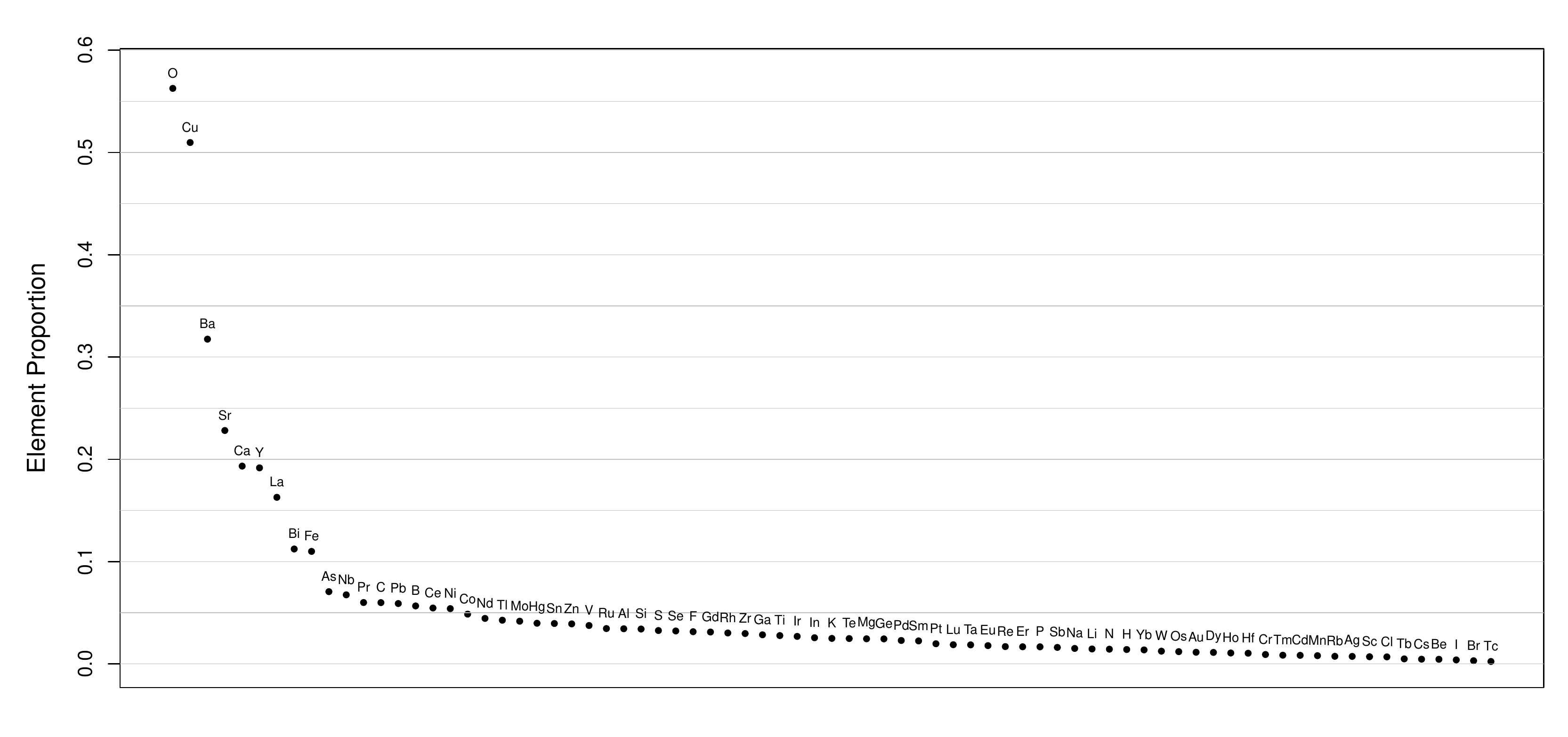}
  \caption{\small This figure shows the proportions of the superconductors that had each element. }\label{fig:element_proportions}
  \end{center}
\end{figure}

Figure (\ref{fig:element_proportions}) shows the proportions of the superconductors that had each element.  For example, Oxygen is present in about 56\% of the superconductors.  Copper, barium, strontium, and calcium are the next most abundant elements.

Iron-based superconductors and cuprates are of particular interest in many research groups so we report some summary statistics in table (\ref{table:summary_for_iron_cuprate}).  Iron is present in approximately 11\% of the superconductors.  The mean $T_c$ of superconductors with iron is $26.9 \pm 21.4$ K.  The non-iron containing superconductors' mean is $35.4 \pm 35.4$ K; the mean and standard deviations happened to be the same.  A t-distribution based 95\% confidence interval suggests that iron containing superconductors' mean $T_c$  is lower than the non-iron's by 7.4 to 9.5 K.  Cuprates comprise approximately 49.5\% of the superconductors.  The cuprates' mean $T_c$ is $59.9 \pm 31.2$ K.  The non-cuprates' mean $T_c$ is $9.5 \pm 10.7$ K.  A t-distribution based 95\% confidence interval indicates that the cuprates' mean $T_c$  is higher than the non-cuprates' mean $T_c$ by 49.8 to 51.0 K.

{\small
\begin{table}
\begin{center}
\begin{tabular}{|c||c|c|c|c|c|c||c|c|}
  \hline
  \           & Size  &  \ Min \ & \ \ Q1 \ \ & Median & \ \  Q3 \ \ & \ Max \ & \ Mean \ & \ \  SD \ \ \\   \hline  \hline
  Iron        & 2339  & 0.02     & 11.3       & 21.7   & 35.5        & 130.0   & 26.9     & 21.4 \\  \hline
  Non-Iron    & 18924 & 0.0002   & 4.8        & 19.6   & 68.0        & 185.0   & 35.4     & 35.4 \\  \hline
  Cuprate     & 10532 & 0.001    & 31.0       & 63.1   & 86.0        & 143     & 59.9     & 31.2 \\  \hline
  Non-Cuprate & 10731 & 0.0002   & 2.5        & 5.7    & 12.2        & 185     & 9.5      & 10.7 \\  \hline
  \hline
\end{tabular}
\caption{This table reports summary statistics on iron-based versus non-iron, and cuprate versus non-cuprate superconductors.  The Size is the total number of observations of the material out of 21,263 materials.  For example, 2,339 out of 21,263 materials contained iron.  The rest of the columns report summary statistics for the observed  critical temperatures (K): min = minimum,  Q1 = first quartile,  Median = median,  Q3 = third quartile, Max = maximum, and SD = standard deviation.
\label{table:summary_for_iron_cuprate}}
\end{center}
\end{table}}

\begin{figure}
\begin{center}
  \includegraphics[width=3.5in]{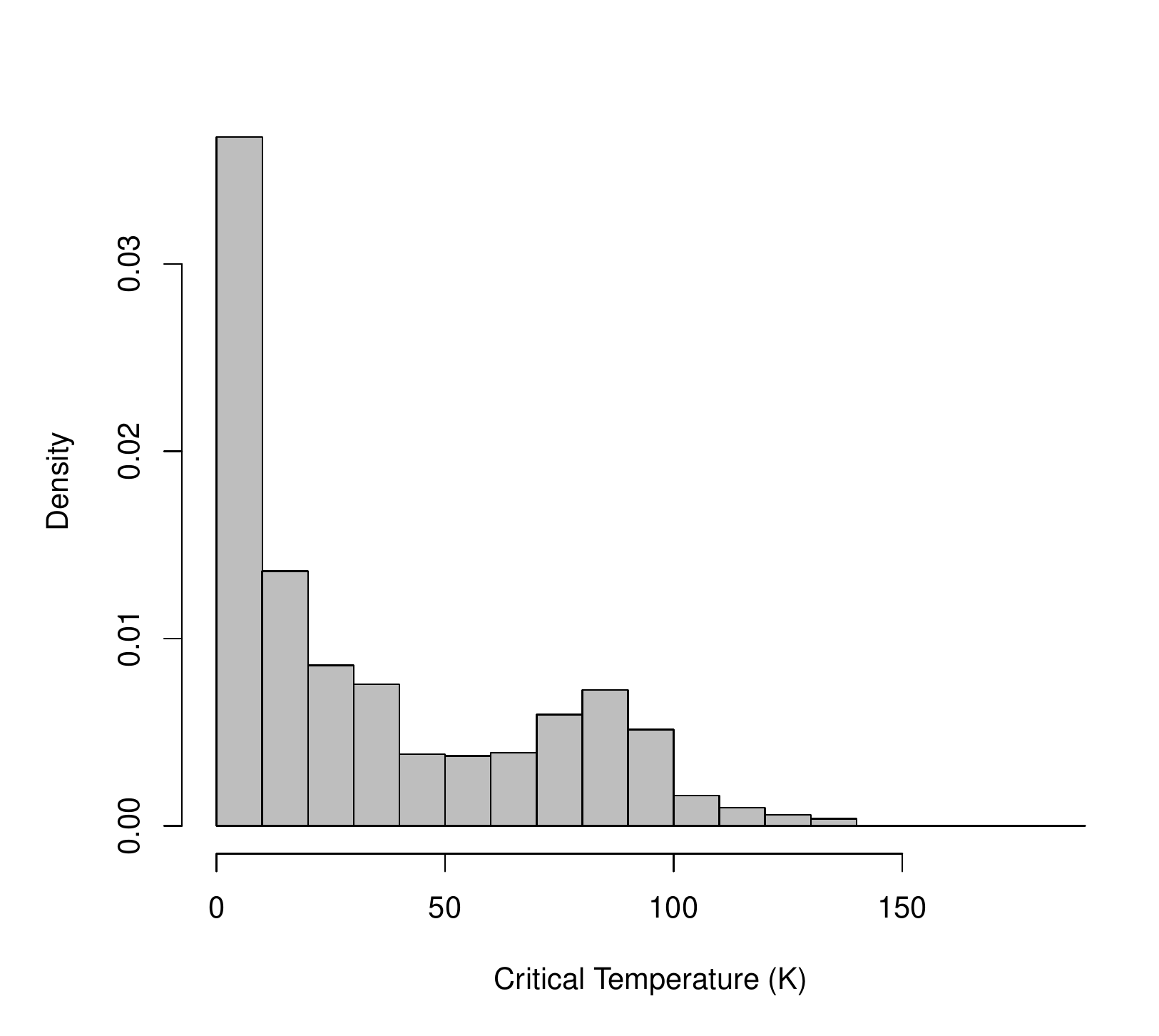}
  \caption{\small This figure shows the distribution of the superconducting critical temperatures (K)  of all 21,263 superconductors. }\label{fig:hist_crit_temp}
  \end{center}
\end{figure}

Figure (\ref{fig:hist_crit_temp}) shows the histogram of $T_c$ values.  The values  are right skewed with a bump around 80 K. Table (\ref{table:critical_temp}) shows the summary statistics for $T_c$ values.

{\small
\begin{table}
\begin{center}
\begin{tabular}{|c|c|c|c|c||c|c|}
  \hline
    \ Min \ & \ \ Q1 \ \ & Median & \ \  Q3 \ \ & \ Max \ & \ Mean \ & \ \  SD \ \ \\   \hline  \hline
        0.00021   & 5.4        & 20    & 63        & 185.0     & 34.4      & 34.2 \\  \hline
  \hline
\end{tabular}
\caption{This table reports the summary statistics for the critical temperatures values (K) of all 21,263 superconductors.  The column headers are the  min = minimum ,  Q1 = first quartile ,  median ,  Q3 = third quartile, Max = maximum, and SD = standard deviation of the superconducting critical temperatures (K).
\label{table:critical_temp}}
\end{center}
\end{table}}

\begin{figure}
\begin{center}
  \includegraphics[width=6in]{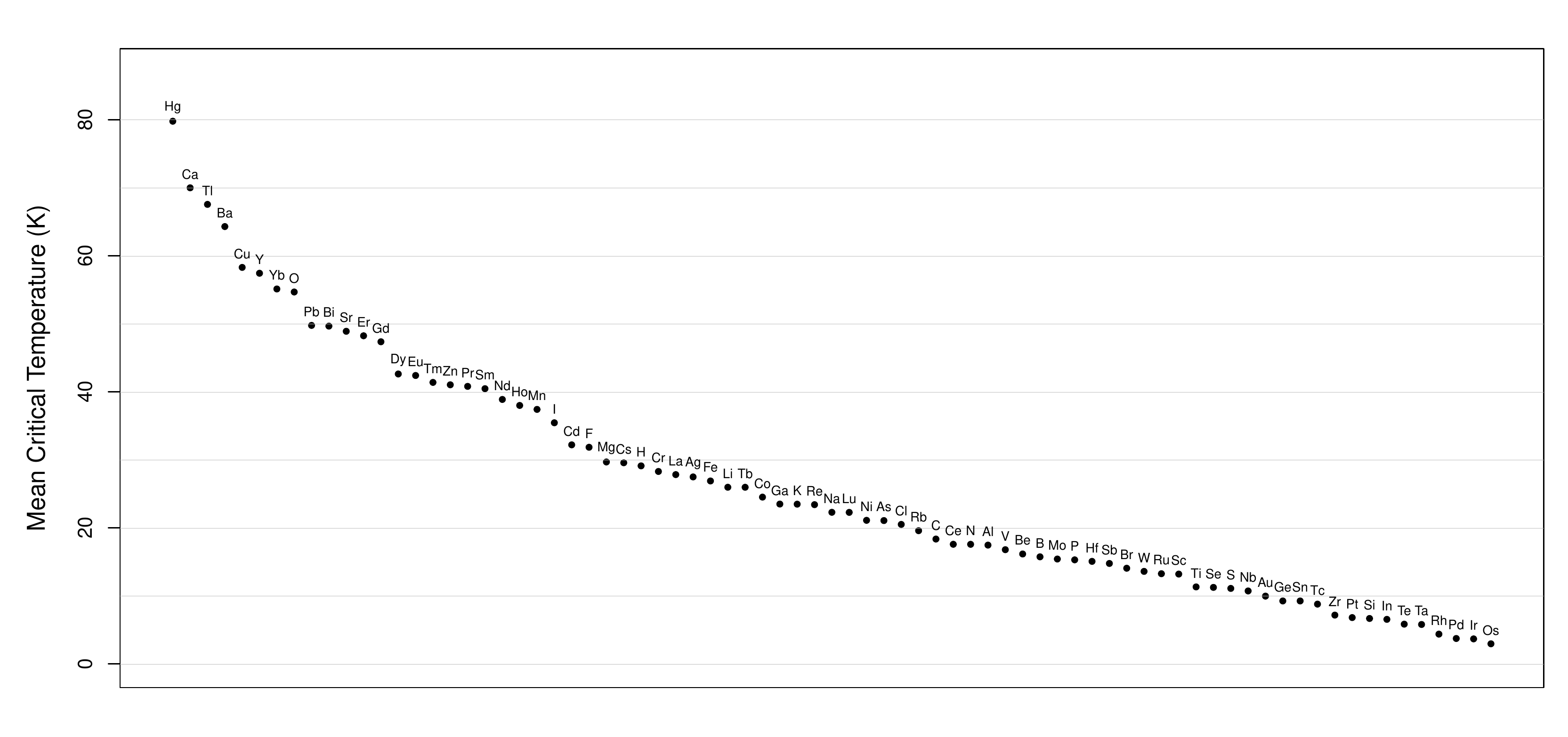}
  \caption{\small This figure shows the mean superconducting critical temperature grouped by elements.  On average, mercury containing materials had the highest superconducting critical temperature followed by calcium and so on.}\label{fig:mean_crit_temp_per_element}
  \end{center}
\end{figure}

\begin{figure}
\begin{center}
  \includegraphics[width=6in]{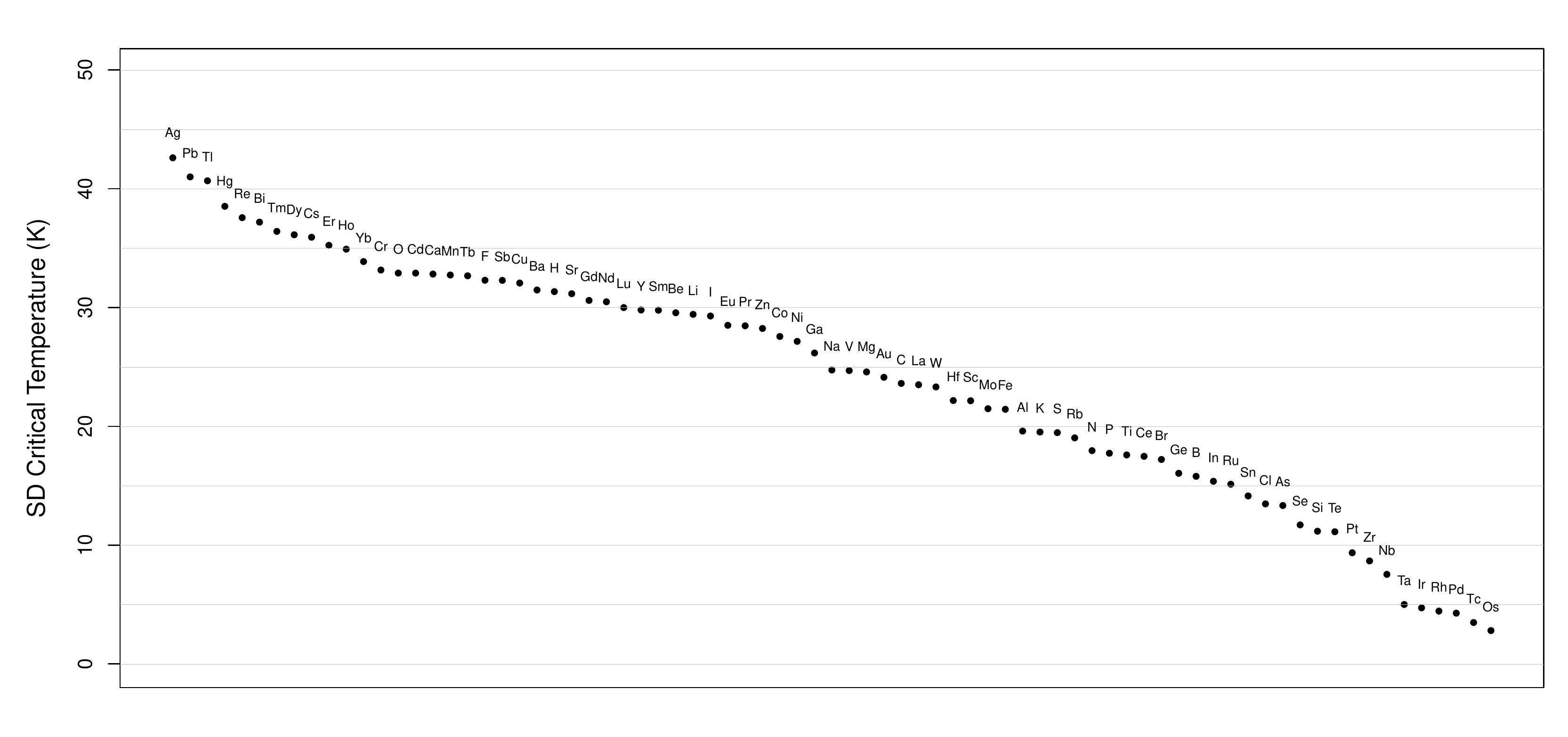}
  \caption{\small This figure shows the standard deviation (SD) of critical temperature grouped by elements.  Silver containing materials had the highest variability followed by lead and so on.}\label{fig:sd_crit_temp_per_element}
  \end{center}
\end{figure}

\begin{figure}
\begin{center}
  \includegraphics[width=6in]{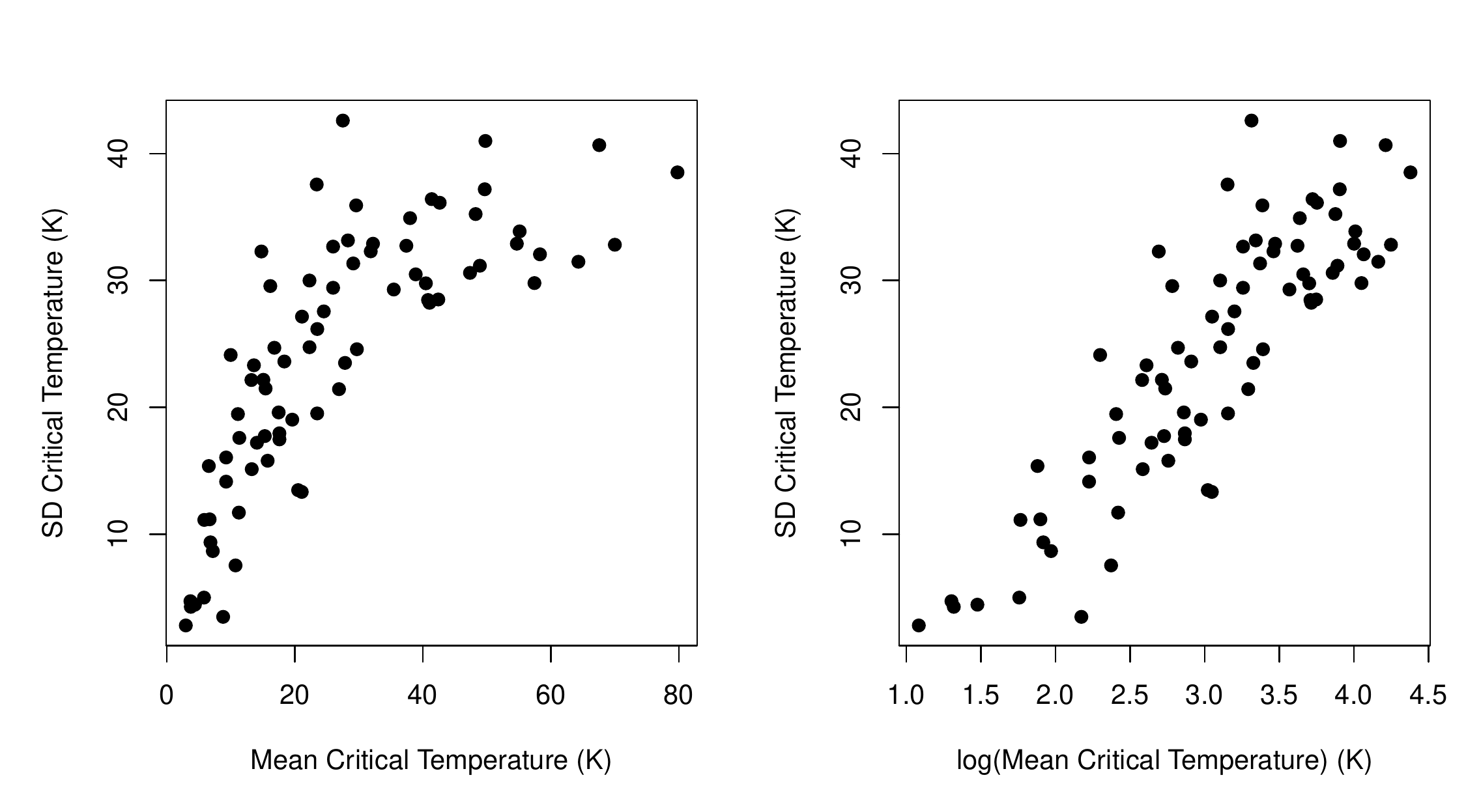}
  \caption{\small The left panel shows the relationship between the mean critical temperature and standard deviation (SD) per element.  The right panel shows the logarithm of the mean critical temperature versus SD.  On average the higher the mean critical temperature, the higher the variability in critical temperature per element.}\label{fig:sd_vs_mean_crit_temp}
  \end{center}
\end{figure}

Figure (\ref{fig:mean_crit_temp_per_element}) shows the mean $T_c$ grouped by elements.  Mercury containing superconductors   have the highest $T_c$ at around 80 K on average.  However, this is not the full story.  Figure (\ref{fig:sd_crit_temp_per_element}) shows the standard deviation of $T_c$ grouped by  elements.  Although mercury containing superconductors have the highest $T_c$ on average, these same materials show the fourth highest variability in $T_c$.  In fact, a plot of the mean $T_c$ versus the standard deviation of $T_c$ in figure (\ref{fig:sd_vs_mean_crit_temp}) shows that on average the higher the mean $T_c$, the higher the variability in $T_c$ per element.

The average absolute value of the correlation among the features is 0.35.  This indicates that the features are highly correlated.  Motivated by this result, we attempted to reduce the dimensionality of the data using principal component analysis (PCA).  However, our PCA analysis did not show any benefits in reducing the dimensionality since a large number of principal components were needed to capture a substantial percentage of the data variation; we abandoned the PCA approach.

\subsection{Model Analysis} \label{subsect:model_analysis}

In this section we discuss the results of the multiple regression model, and the XGBoost model. We tried a few classical models including multiple regression with interactions, principal component regression, and partial least squares but none of these make any substantial improvements to the XGBoost model.  We also tried  random forests but they were too slow to tune given the data size and the number of features.  Scalability and speed are important advantages of using XGBoost over random forests; See \cite{chen2016}.

The prediction performance of the models are compared by using out-of-sample rmse. The out-of-sample rmse is estimated by the following cross validation procedure: \vspace{2 mm}

\underline{\textbf{Out-Of-Sample RMSE Estimation Procedure:}} \label{algo:cv}
\begin{enumerate}
  \item At random, divide the data into $2/3$ train data and $1/3$ test data.
  \item Fit the model using the train data.
  \item Predict $T_c$ of the test data.
  \item Obtain an estimate of the out-of-sample mean-squared-error (mse) by using the predictions from the last step and the observed $T_c$ values  in the test data:
      $$\text{out-of-sample mse} = \text{Average of (observed - predicted)}^2$$
  \item Repeat steps 1 through 4, 25 times to collect 25 out-of-sample mse's.
  \item Take the mean of the 25 collected out-of-sample mse's and report the square root of this average as the final estimate of the out-of-sample rmse.
\end{enumerate}

\subsubsection{The Multiple Regression Model} \label{subsubsect:multiple_regression}
\begin{figure}
\begin{center}
  \includegraphics[width=3.5in]{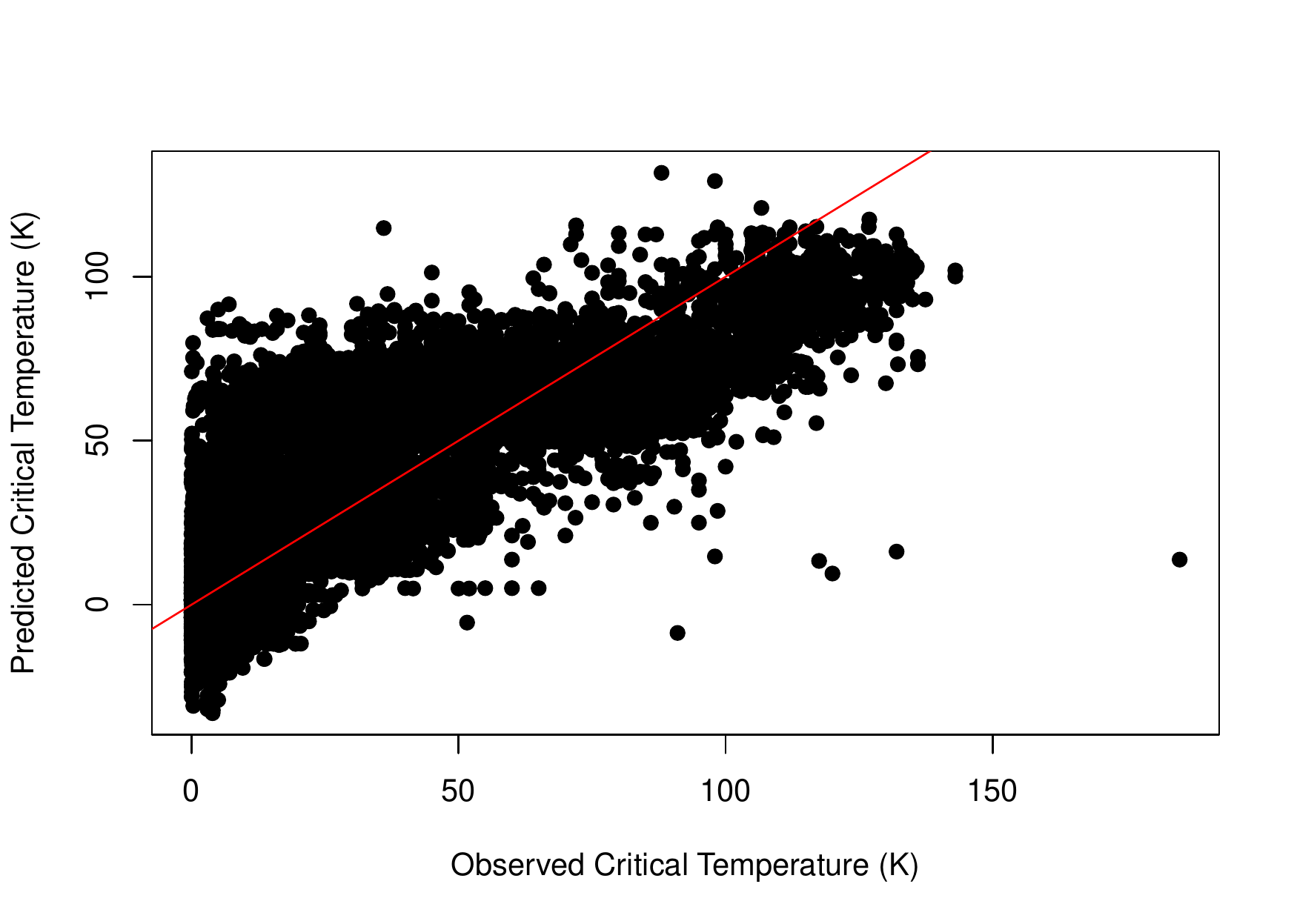}
  \caption{\small This plot shows  the predicted superconducting critical temperatures (K) versus the observed superconducting critical temperatures (K) based on the multiple regression model.  The out-of-sample rmse is about $17.6$ K.  The out-of-sample $R^2$ is about 0.74. }\label{fig:linear_model_predicted_tc_vs_observed_tc}
  \end{center}
\end{figure}

The multiple regression model's out-of-sample rmse estimated by the procedure above is about 17.6 K.  The out-of-sample $R^2$ is about $0.74$.   Figure (\ref{fig:linear_model_predicted_tc_vs_observed_tc}) shows the predicted $T_c$ versus the observed $T_c$ when we use all the data to fit the model.  The line has an intercept of zero and a slope of 1.  The plot indicates that the multiple regression model under-predicts $T_c$ of high temperature superconductors since many predicted points are below the line for the high temperature superconductors.  The model over-predicts low  temperature superconductors' $T_c$. The multiple regression model  simply serves as a benchmark model and should not be used for prediction.  There would be no use in predicting $T_c$ using a sophisticated model such as XGBoost, if a commonly used  multiple regression model does a good job.  Here, the XGBoost model vastly improves the prediction accuracy.

\subsubsection{The XGBoost Model} \label{subsubsect:gbm}

Before we go on, we give a brief description of  XGBoost set up.  XGBoost is described in detail in \cite{chen2016}. A readable summary is given at \url{https://xgboost.readthedocs.io/en/latest/model.html}. \cite{elements_of_stat_learning} and \cite{modern_mul_stat} give general overviews on boosting as well.

The functional form of XGBoost is:
$$\hat{y}_i = \sum_{k = 1}^{K}f_k(x_i),$$
where $x_i$ is the $i$th input feature vector, $\hat{y}_i$ is the predicted response, and $f_1,\ldots,f_K$ is a sequence of trees.  The $t$-th tree $f_t$ is added by minimizing the following objective function:
\begin{equation} \label{equ:loss_function}
\text{Objective with respect to }f_t = \sum_{i=1}^{n}L\big(\underbrace{y_i}_{observed},\ \underbrace{\hat{y}^{(t-1)}_i + f_t(x_i)}_{predicted} \big) + \Omega(f_t),
\end{equation}
where $L$ is the desired loss function, $n$ is the total sample size, $y_i$'s are the response values, $\hat{y}^{(t-1)}_i$ is the $i$th predicted responses at the $t-1$ step, and $\Omega$ is a penalty function.  The form of $\Omega$ is:
\begin{equation} \label{equ:constraint}
\Omega(f) = \gamma T + (1/2)\lambda \sum_{j=1}^{T} w_j^2,
\end{equation}
where $T$ is the number of leaves in each tree, $w_j$'s are the leaf weights, and $\lambda$ and $\gamma$ are regularization parameters.  The goal here is to add a new tree $f_t$ to the overall ensemble of trees to minimizes the loss between the observed and the predicted in equation (\ref{equ:loss_function}), while preventing over-fitting by satisfying the penalty in equation (\ref{equ:constraint}).  The addition of this penalty function to \emph{each} tree in (\ref{equ:constraint}) is one major XGBoost differentiator from the established method by \cite{friedman2001}.  The penalty function appears to make a big difference in practice; see \cite{chen2016}.  Besides the clever penalty function, \cite{chen2016} implement numerous computational tricks to make their software scalable and very fast.

In addition to the penalty function, there are a number of tuning parameters that could reduce over-fitting and enhance the model's prediction performance; They are mainly: (1) column subsampling which means only a fraction of the features are chosen at random at each stage of adding a new tree, (2) a learning parameter $0 < \eta < 1$ which scales the contribution of each new tree, (3) subsample ratio which means that XGBoost only uses a small percentage of the data  to grow a new tree, (4) maximum depth of a tree, and (5) minimum child weight which is the minimum number of data points needed to be in each node.

To tune XGBoost, we first split the data at random  to 2/3 train and 1/3 test data.   Next, we create a grid  - a grid contains all the possible combination of tuning parameters - with $\eta = 0.010, 0.015, 0.020$, column subsampling = 0.25, 0.5, 0.75, subsample ratio = 0.5, minimum node size = 1, 10, and maximum depth of a tree = 15, 16, ..., 24, 25.  The total gird size is 198.  This means that we need 198 different XGBoost models.  For each model, 750 trees are grown.  The rest of the XGBoost parameters are set to the default values.  (This was not our only grid; we had done some experimentations with various grids  before we decided to use this  grid.)  Finally, we evaluate the prediction accuracy of each model based on rmse  at each tree = 1, 2, ..., 749, 750.

The best model (with the lowest out-of-sample rmse) turn out to be: $\eta = 0.02$, maximum depth $=16$, minimum child weight $= 1$, column subsampling $=0.50$, and a tree size of 374.  To obtain the final out-of-sample rmse and $R^2$, we follow the 6 step procedure outlined at the begining of section (\ref{subsect:model_analysis}).  The procedure yield an out-of-sample rmse of $9.5$ K, and a out-of-sample $R^2$ of 0.92.  The out-of-sample rmse of $9.5$ K has a very important interpretation: On average, the tuned XGBoost model will be off by about $9.5$ K when predicting $T_c$.

\begin{figure}
\begin{center}
  \includegraphics[width=3.5in]{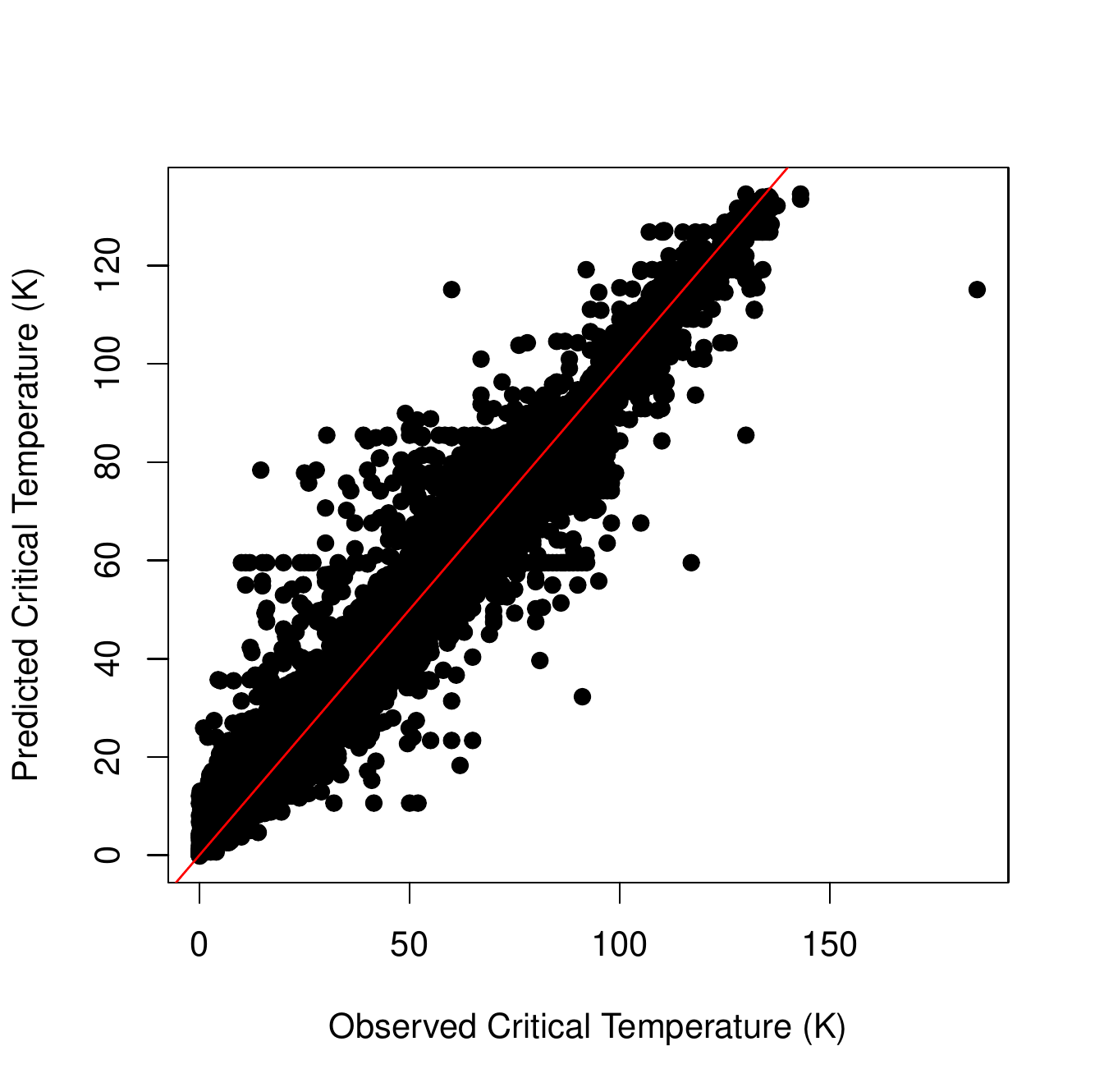}
  \caption{\small This plot shows  the predicted critical temperatures versus observed critical temperatures (K) based on the XGBoost model.  The out-of-sample rmse is $9.4$ K.  The out-of-sample $R^2$ is 0.92. }\label{fig:xgboost_out_of_sample_predicted_tc_vs_observed_tc}
  \end{center}
\end{figure}

Figure (\ref{fig:xgboost_out_of_sample_predicted_tc_vs_observed_tc}) shows the  predicted $T_c$ versus the observed $T_c$. Except for lower observed $T_c$ vlues, no severe bias is discernable.  There are are a number outliers visible.

\newpage
\subsubsection{Feature Importance} \label{subsubsect:feature_importance}

Feature importance in XGBoost is measured by gain.  The gain for a feature is defined as follows: Whenever a tree is split on a feature, the improvement in the objective function is recorded. The gain for the feature is then:
$$\text{The Gain for the Feature} = \frac{\text{Sum of the Gains for the Feature}}{\text{Sum of the Gains for All the Features}}.$$
Features with higher gain are more important.

Table (\ref{table:var_importance}) shows the top 20 most important features. Features extracted based on thermal conductivity, atomic radius, valence,  electron affinity, and atomic mass  appear to be the most important features.  Also observe that features defined based on thermal conductivity, valence, electron affinity, and atomic mass appear most often on the list.  This may suggest that these  properties could be more important than other properties in predicting $T_c$.

{\small
\begin{table}
\begin{center}
\begin{tabular}{|l|c|}
  \hline
\textbf{Feature}	&	\textbf{Gain}	\\ \hline \hline
range\_ThermalConductivity	&	0.295	\\ \hline
wtd\_std\_ThermalConductivity	&	0.084	\\ \hline
range\_atomic\_radius	        &	0.072	\\ \hline
wtd\_gmean\_ThermalConductivity	&	0.047	\\ \hline
std\_ThermalConductivity	&	0.042	\\ \hline
wtd\_entropy\_Valence	        &	0.038	\\ \hline
wtd\_std\_ElectronAffinity	&	0.036	\\ \hline
wtd\_entropy\_atomic\_mass	&	0.025	\\ \hline
wtd\_mean\_Valence	        &	0.022	\\ \hline
wtd\_gmean\_ElectronAffinity	&	0.021	\\ \hline
wtd\_range\_ElectronAffinity	&	0.016	\\ \hline
wtd\_mean\_ThermalConductivity	&	0.015	\\ \hline
wtd\_gmean\_Valence	        &	0.014	\\ \hline
std\_atomic\_mass	        &	0.013	\\ \hline
std\_Density	                &	0.010	\\ \hline
wtd\_entropy\_ThermalConductivity	&	0.010	\\ \hline
wtd\_range\_ThermalConductivity	&	0.010	\\ \hline
wtd\_mean\_atomic\_mass	        &	0.009	\\ \hline
wtd\_std\_atomic\_mass	        &	0.009	\\ \hline
gmean\_Density	                &	0.009	\\ \hline
  \hline
\end{tabular}
\caption{ This figure shows the top 20 most important features based on the XGBoost gain.  Here: wtd = weighted, gmean = geometric mean, std = standard deviation.
\label{table:var_importance}}
\end{center}
\end{table}}

\clearpage

\section{Prediction Software}

We have put the code for prediction at \url{https://github.com/khamidieh/predict_tc}.  The software is created using R Statistical programming language, \cite{r}.    The data could also be directly downloaded from our github site.

\begin{figure}
\begin{center}
  \includegraphics[width=3in]{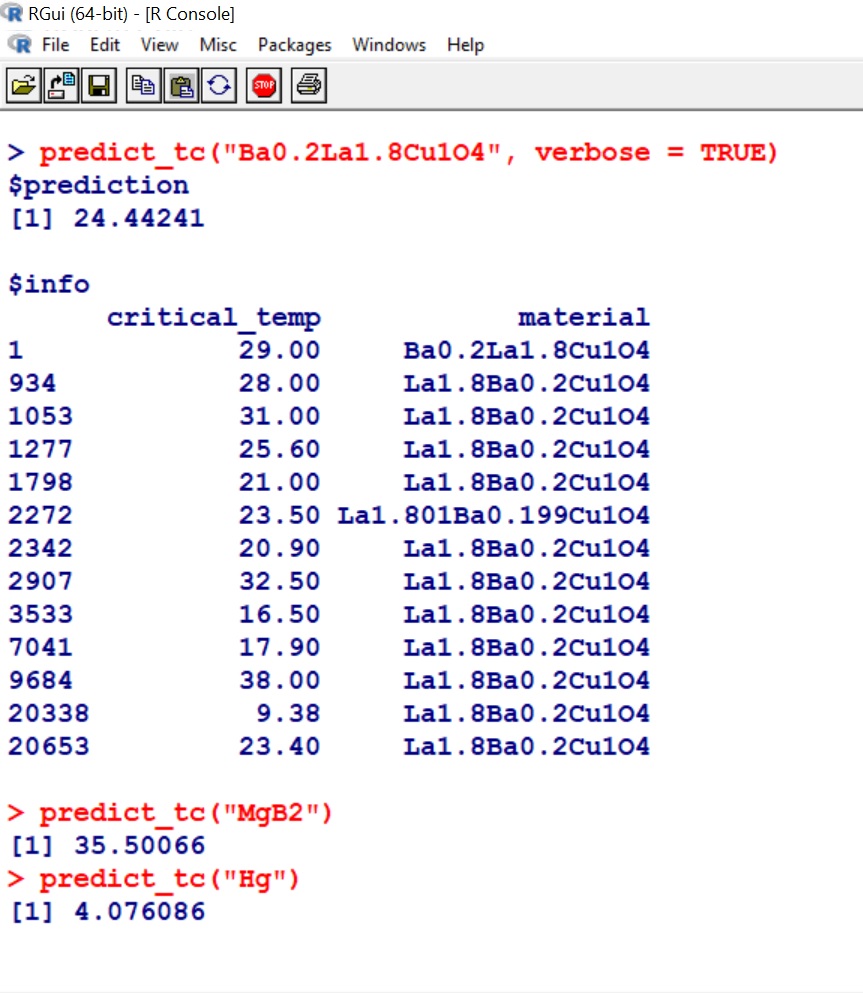}
  \caption{\small This figure shows the software prediction results for \chem{Ba_{0.2}La_{1.8}CuO_{4}}, \chem{MgB{2}}, and  \chem{Hg}. }\label{fig:softdemo_1_xgb}
  \end{center}
\end{figure}

We demonstrate some examples using the software.  Figure (\ref{fig:softdemo_1_xgb}) shows the predictions for three materials: \chem{Ba_{0.2}La_{1.8}CuO_{4}}, \chem{MgB_{2}}, and \chem{Hg}.  The ``verbose" option uses the cosine similarity measure to  pull data with similar chemical formulas.  The multiple entries for \chem{Ba_{0.2}La_{1.8}CuO_{4}} are obtained.  The default value for verbose is false so no superconductors similar to \chem{MgB_{2}} and \chem{Hg} are shown.

\begin{figure}
\begin{center}
  \includegraphics[width=2.5in]{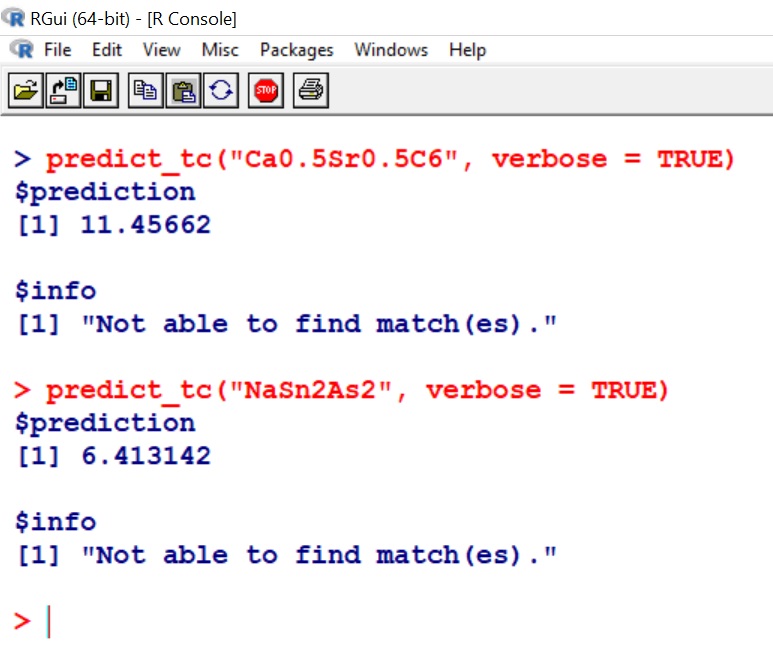}
  \caption{\small This figure shows the software prediction results for \chem{Ca_{0.5}Sr_{0.5}C_{6}} and \chem{NaSn_{2}As_{2}} which have reported critical temperatures of 3 K and 1.3 K respectively. }\label{fig:softdemo_2_xgb}
  \end{center}
\end{figure}

We had obtained the data on July 24, 2017. We like to see what sort of predictions we could obtain for some new superconductors reported since.  \cite{saki} report a $T_c$ of around 3 K  for \chem{Ca_{0.5}Sr_{0.5}C_{6}}. \cite{goto} report a $T_c$ of 1.3 K for  \chem{NaSn_{2}As_{2}}.  Figure (\ref{fig:softdemo_2_xgb}) shows the prediction results.  The XGBoost model over-predicts but it is within the $\pm 9.5$ K out-of-sample rmse.  The message ``Not able to find match(es)" indicates that nothing in the training data is similar to these two new superconductors.  We should not expect good predictions for completely new superconductors.

\begin{figure}
\begin{center}
  \includegraphics[width=2.5in]{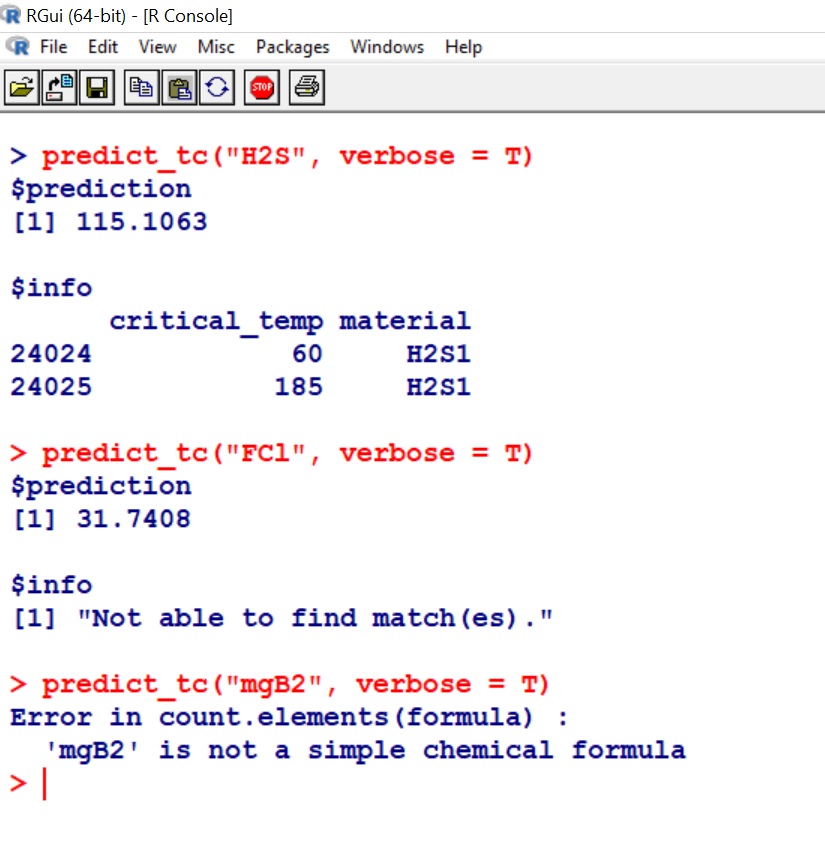}
  \caption{\small This figure shows the software prediction results for \chem{H_{2}S}, and (non-sense) \chem{FCl}, and misspelled formula \chem{mgB{2}}. }\label{fig:softdemo_3_xgb}
  \end{center}
\end{figure}

Figure (\ref{fig:softdemo_3_xgb}) shows what can go wrong when the XGBoost model predicts badly or when the inputs do not make sense.  The prediction for \chem{H_{2}S}, which has a $T_c$ of 203 K under extremely high pressures, is way off.  (Note that \chem{H_{2}S} with $T_c$ of 203 is not in the train data.)  This is perhaps expected since there is no feature that captures the dependence of $T_c$ on pressure.  The model gives a prediction for \chem{FCl} but this is a non-sense;  The prediction model can't check for the existence of solids.  The model gives an error message for \chem{mgB{2}} since it does not recognize \chem{mg} with the lower case m as an element.

Next, we predict $T_c$ for materials identified by \cite{Stanev} as potential superconductors.  The results are shown in table (\ref{table:potential_superconductors}).  None of the superconductors in table (\ref{table:potential_superconductors}) are found to be (cosine) similar to the superconductors in our train data.

{\small
\begin{table}
\begin{center}
\begin{tabular}{|l|c|}
  \hline
\textbf{Material}	                    &	\textbf{Predicted} $T_c$ (K)	\\ \hline \hline
\chem{CsBe(AsO_{4})}	                &	13.7	\\ \hline
\chem{RbAsO_{2}}	                    &	8.0	    \\ \hline
\chem{KSbO_{2}}	                        &	10.2	\\ \hline
\chem{RbSbO_{2}}                    	&	11.8	\\ \hline
\chem{CsSbO_{2}}	                    &	10.1	\\ \hline
\chem{AgCrO_{2}}	                    &	53.3	\\ \hline
\chem{K_{0.8}(Li_{0.2}Sn_{0.76})O_{2}}	&	18.6	\\ \hline
\chem{Cs(MoZn)(O_{3}F_{3})}	            &	20.5	\\ \hline
\chem{Na_{3}Cd_{2}(IrO_{6})}	        &	17.4	\\ \hline
\chem{Sr_{3}Cd(PtO_{6})}	            &	12.8	\\ \hline
\chem{Sr_{3}Zn(PtO_{6})}	            &	12.4	\\ \hline
\chem{(Ba_{5}Br_{2})Ru_{2}O_{9}}	    &	17.0	\\ \hline
\chem{Ba_{4}(AgO_{2})(AuO_{4})}     	&	56.7	\\ \hline
\chem{Sr_{5}(AuO_{4})_{2}}          	&	17.8	\\ \hline
\chem{RbSeO_{2}F}                   	&	16.7	\\ \hline
\chem{CsSeO_{2}F}	                    &	20.4	\\ \hline
\chem{KTeO_{2}F}	                    &	13.0	\\ \hline
\chem{Na_{2}K_{4}(Tl_{2}O_{6})}     	&	32.8	\\ \hline
\chem{Na_{3}Ni_{2}BiO_{6}}	            &	17.1	\\ \hline
\chem{Na_{3}Ca_{2}BiO_{6}}          	&	27.3	\\ \hline
\chem{CsCd(BO_{3})}	                    &	22.3	\\ \hline
\chem{K_{2}Cd(SiO_{4})}             	&	17.7	\\ \hline
\chem{Rb_{2}Cd(SiO_{4})}	            &	17.4	\\ \hline
\chem{K_{2}Zn(SiO_{4})}             	&	19.6	\\ \hline
\chem{K_{2}Zn(Si_{2}O_{6})}	            &	12.2	\\ \hline
\chem{K_{2}Zn(GeO_{4})}	                &	17.6	\\ \hline
\chem{(K_{0.6}Na_{1.4})Zn(GeO_{4})}	    &	25.6	\\ \hline
\chem{K_{2}Zn(Ge_{2}O_{6})}	            &	10.4	\\ \hline
\chem{Na_{6}Ca_{3}(Ge_{2}O_{6})_{3}}	&	12.1	\\ \hline
\chem{Cs_{3}(AlGe_{2}O_{7})}	        &	14.8	\\ \hline
\chem{K_{4}Ba(Ge_{3}O_{9})}         	&	15.1	\\ \hline
\chem{K_{16}Sr_{4}(Ge_{3}O_{9})_{4}}	&	13.5	\\ \hline
\chem{K_{3}Tb[Ge_{3}O_{8}(OH)_{2}]}	    &	11.2	\\ \hline
\chem{K_{3}Eu[Ge_{3}O_{8}(OH)_{2}]}	    &	11.3	\\ \hline
\chem{KBa_{6}Zn_{4}(Ga_{7}O_{21})}	    &	30.1	\\ \hline
  \hline
\end{tabular}
\caption{ This table shows $T_c$ predictions for a list of potential superconductors identified by \cite{Stanev}.
\label{table:potential_superconductors}}
\end{center}
\end{table}}


\section{Conclusion}

We have shown that a statistical model using only the superconductors' chemical formula can predict $T_c$ reasonably well. We have also made the software and the data easily available.   There are practical uses for our model: (1) Researchers interested in finding high temperature superconductors may use the model to narrow their search, and (2) researchers could use the cleaned data along with new data (such as pressure or crystal structure) to make better models.

\section{Acknowledgements}

We like to thank Dr. Allan Macdonald, professor of physics at the University of Texas at Austin, for many useful suggestions.

\newpage
\section{Bibliography}

\bibliography{bib}

\newpage
\appendix

\section{Mathematica ElementData} \label{appendix:data_element}

Below is the list of sources Mathematica has used to obtained the element property data.  It is directly copied from: \newline
 \url{http://reference.wolfram.com/language/note/ElementDataSourceInformation.html}.

\begin{itemize}
  \item Atomic Mass Data Center. ``NUBASE." 2003. \url{http://amdc.in2p3.fr/web/nubase_en.html}
  \item Cardarelli, F. Materials Handbook: A Concise Desktop Reference. Springer, 2000.
  \item Lide, D. R. (Ed.). CRC Handbook of Chemistry and Physics. 87th ed. CRC Press, 2006.
  \item Speight, J. Lange's Handbook of Chemistry. McGraw-Hill, 2004.
  \item United Kingdom National Physical Laboratory. ``Kaye and Laby Tables of Physical and Chemical Constants." \url{http://www.kayelaby.npl.co.uk/}
  \item United States National Institute of Standards and Technology. ``Atomic Weights and Isotopic Compositions Elements." \newline
        \url{https://www.nist.gov/pml/atomic-weights-and-isotopic-compositions-relative-atomic-masses}
  \item United States National Institute of Standards and Technology. ``NIST Chemistry Webbook."  \url{http://webbook.nist.gov/chemistry/}
  \item Winter, M. ''WebElements." 2007. \url{https://www.webelements.com/}
\end{itemize}

\end{document}